# Exploiting Multiple-Antenna Techniques for Non-Orthogonal Multiple Access

Xiaoming Chen, *Senior Member, IEEE*, Zhaoyang Zhang, *Member, IEEE*, Caijun Zhong, *Senior Member, IEEE*, and Derrick Wing Kwan Ng, *Member, IEEE*

## Abstract

This paper aims to provide a comprehensive solution for the design, analysis, and optimization of a multiple-antenna non-orthogonal multiple access (NOMA) system for multiuser downlink communication with both time duplex division (TDD) and frequency duplex division (FDD) modes. First, we design a new framework for multiple-antenna NOMA, including user clustering, channel state information (CSI) acquisition, superposition coding, transmit beamforming, and successive interference cancellation (SIC). Then, we analyze the performance of the considered system, and derive exact closed-form expressions for average transmission rates in terms of transmit power, CSI accuracy, transmission mode, and channel conditions. For further enhancing the system performance, we optimize three key parameters, i.e., transmit power, feedback bits, and transmission mode. Especially, we propose a low-complexity joint optimization scheme, so as to fully exploit the potential of multiple-antenna techniques in NOMA. Moreover, through asymptotic analysis, we reveal the impact of system parameters on average transmission rates, and hence present some guidelines on the design of multiple-antenna NOMA. Finally, simulation results validate our theoretical analysis, and show that a substantial performance gain can be obtained over traditional orthogonal multiple access (OMA) technology under practical conditions.

Xiaoming Chen (chen_xiaoming@zju.edu.cn), Zhaoyang Zhang (ning_ming@zju.edu.cn), and Caijun Zhong (caijunzhong@zju.edu.cn) are with the College of Information Science and Electronic Engineering, Zhejiang University, Hangzhou 310027, China. Derrick Wing Kwan Ng (w.k.ng@unsw.edu.au) is with the school of Electrical Engineering and Telecommunications, the University of New South Wales, NSW, Australia.

I. INTRODUCTION

Current wireless communication systems in general adopt various types of orthogonal multiple access (OMA) technologies for serving multiple users, such as time division multiple access (TDMA), frequency division multiple access (FDMA), and code division multiple access (CDMA), where one resource block is exclusively allocated to one mobile user (MU) to avoid possible multiuser interference. In practice, the OMA technologies are relatively easy to implement, albeit at the cost of low spectral efficiency. Recently, with the rapid development of mobile internet and proliferation of mobile devices, it is expected that future wireless communication systems should be able to support massive connectivity, which is an extremely challenging task for the OMA technologies with limited radio resources. Responding to this, non-orthogonal multiple access (NOMA) has been recently proposed as a promising access technology for the fifth-generation (5G) mobile communication systems, due to its potential in achieving high spectral efficiency [1]–[3].

The principle of NOMA is to exploit the power domain to simultaneously serve multiple MUs utilizing the same radio resources [4], [5], with the aid of sophisticated successive interference cancellation (SIC) receivers [6], [7]. Despite the adoption of SIC, inter-user interference still exists except for the MU with the strongest channel gain, which limits the overall system performance [8]. To address this issue, power allocation has been considered as an effective method to harness multiuser interference [9], [10]. Since the overall performance is limited by the MUs with weak channel conditions, it is intuitive to allocate more power to the weak MUs and less power to the strong MU in order to enhance the effective channel gain and minimize the interference to the weak MUs [11]. For the specific two-user case, the optimal power allocation scheme was studied in [12], and [13] proposed two sub-optimal power allocation schemes exploiting the Karush-Kuhn-Tucker (KKT) conditions, while the issue of quality of service (QoS) requirements of NONA systems was investigated in [14]. For the case with arbitrary number of users, the computational complexity of performing SIC increases substantially and the design of the optimal power allocation becomes intractable. To facilitate an effective system design, clustering and user pairing have been proposed [15], [16]. Generally speaking, multiple MUs with distinctive channel gains are selected to form a cluster, in which SIC is conducted to mitigate the interference [17], [18]. In general, a small cluster consisting a small number of MUs implies low complexity of SIC, but leads to high inter-cluster interference. Thus, it makes sense to dynamically adjust the size of a cluster according to performance requirements and system parameters, so as to achieve a balance between implementation complexity and interference mitigation [19]. However, dynamic clustering is not able to reduce the inter-cluster interference, indicating the necessity of carrying out dynamic clustering in combination with efficient interference mitigation schemes.

It is well known that the multiple-antenna technology is a powerful interference mitigation scheme [20]–[22], hence, can be naturally applied to NOMA systems [23], [24]. In [25], the authors proposed a beamforming scheme for combating inter-cluster and intra-cluster interference in a NOMA downlink, where the base station (BS) was equipped with multiple antennas and the MUs have a single antenna each. A more general setup was considered in [26], where both the BS and the MUs are multiple-antenna devices. By exploiting multiple antennas at the BS and the MUs, a signal alignment scheme was proposed to mitigate both the intra-cluster and inter-cluster interference. It is worth pointing out that the implementation of the two above schemes requires full channel state information (CSI) at the BS, which is usually difficult and costly in practice. To circumvent the difficulty in CSI acquisition, random beamforming was adopted in [27], which inevitably leads to performance loss. Alternatively, the work in [28] suggested to employ zero-forcing (ZF) detection at the multiple-antenna MUs for inter-cluster interference cancelation. However, the ZF scheme requires that the number of antennas at each MU is greater than the number of antennas at the BS, which is in general impractical.

To effectively realize the potential benefits of multiple-antenna techniques, the amount and quality of CSI available at the BS plays a key role. In practice, the CSI can be obtained in several different ways. For instance, in time duplex division (TDD) systems, the BS can obtain the downlink CSI through estimating the CSI of uplink by leveraging the channel reciprocity. While in frequency duplex division (FDD) systems, the downlink CSI is usually first estimated and quantized at the MUs, and then is conveyed back to the BS via a feedback link. For both practical TDD and FDD systems, the BS has access to only partial CSI. As a result, there will be residual inter-cluster and intra-cluster interference. To the best of the authors' knowledge, previous works only consider two extreme cases with full CSI or no CSI, the design, analysis and optimization of multiple-antenna NOMA systems with partial CSI remains an uncharted area.

Motivated by this, we present a comprehensive study on the impact of partial CSI on the design, analysis, and optimization of multiple-antenna NOMA downlink communication systems. Specifically, we consider heterogeneous downlink channels, and the BS equipped with arbitrarily multiple antennas has different CSI accuracies about the downlink channels. The major contributions of this paper are summarized as follows:

1) We design a general framework for multiple-antenna NOMA downlink communications including user clustering, CSI estimation, superposition coding, transmit beamforming, and SIC. In particular, the proposed framework is applicable in both TDD and FDD modes.

2) We analyze the performance of the proposed multiple-antenna NOMA, and derive exact expressions for the average transmission rates of each MU in an arbitrary cluster. The average transmission rate is a function of transmit power, CSI accuracy, transmission mode, and channel conditions.

3) We optimize three key parameters of multiple-antenna NOMA, namely, transmit power, feedback bits, and transmission mode. In particular, we present closed-form expressions for the power allocation and feedback distribution. For mode selection, we show that the mode of two MUs in a cluster is optimal in practical cases with moderate and high CSI accuracy, which provides theoretical justification for the two-user setup in the previous works [9]–[14], [23]–[28], [30], [31]. Finally, a low complexity joint optimization scheme of transmit power, feedback bits, and transmission mode is proposed.

4) Through asymptotic analysis of average transmission rates, several key insights are obtained.

    a) Imperfect CSI results in residual inter-cluster interference at MUs. Thus, there exists a performance gap between practical NOMA with imperfect CSI and ideal NOMA with perfect CSI. The performance gap is an increasing function of transmit power of information signal and a decreasing function of CSI accuracy. In order to maintain a constant performance, transmit energy of pilot sequence for channel estimation in TDD mode and spatial resolution in FDD mode should be increased as transmit

power of information signal grows.

b) From the perspective of maximizing the sum rate, arranging all MUs in one cluster is optimal if there is no CSI at the BS, while the best option is to arrange one MU in each cluster if there is perfect CSI at the BS.

c) In the interference-limited scenario, the average transmission rate for the 1st MU with the strongest channel gain in each cluster increases linearly proportionally to the number of feedback bits. Under the noise-limited condition, the average transmission rate is independent of CSI accuracy.

d) In the interference-limited case, equal power allocation among all MUs asymptotically approaches the optimal performance.

The rest of this paper is organized as follows: Section II gives a brief introduction of the considered NOMA downlink communication system, and designs the corresponding multiple-antenna transmission framework. Section III first analyzes the average transmission rates in presence of imperfect CSI, and then proposes three performance optimization schemes. Section IV derives the average transmission rates in two extreme cases through asymptotic analysis, and presents some system design guidelines. Section V provides simulation results to validate the effectiveness of the proposed schemes. Finally, Section VI concludes the whole paper.

*Notations*: We use bold upper (lower) letters to denote matrices (column vectors), $(\cdot)^H$ to denote conjugate transpose, $\mathrm{E}[\cdot]$ to denote expectation, $\|\cdot\|$ to denote the $L_2$-norm of a vector, $|\cdot|$ to denote the absolute value, $\stackrel{d}{=}$ to denote the equality in distribution, $\lfloor x \rfloor$ to denote the maximum integer not larger than $x$, and $\mathbb{C}$ to denote the set of complex number. The acronym i.i.d. means "independent and identically distributed", pdf means "probability density function", and cdf means "cumulative distribution function".

## II. System Model and Framework Design

### A. User Clustering

Consider a downlink communication scenario in a single-cell system, where a base station (BS) broadcasts messages to multiple MUs, cf. Fig. 1. Note that the BS is equipped with $M$ antennas, while the MUs have a single antenna each due to the size limitation. To strike a balance between the system performance and computational complexity in NOMA systems, it is necessary to carry out user clustering. In particular, user clustering can be designed from different perspectives. For instance, a signal-to-interference-plus-noise ratio (SINR) maximization user clustering scheme was adopted in [32] and quasi-orthogonal MUs were selected to form a cluster in [33]. Intuitively, these schemes perform user clustering by the exhaustive search method, resulting in high implementation complexity. In this paper, we design a simple user clustering scheme based on the information of spatial direction[1]. Specifically, the MUs in the same direction but with distinctive propagation distances are arranged into a cluster. On one hand, the same direction of the MUs in a cluster allows the use of a single beam to nearly align all MUs in such a cluster, thereby facilitating the mitigation of the inter-cluster interference and the enhancement of the effective channel gain. On the other hand, a large gap of propagation distances avoids severe inter-user interference and enables a more accurate SIC at the MUs [34]–[36]. If two MUs are close to each other with almost equal channel gains, it is possible to assign them in different clusters by improving the spatial resolution via increasing the number of spatial beams and the number of BS antennas. Without loss of generality, we assume that the MUs are grouped into $N$ clusters with $K$ MUs in each cluster. To facilitate the following presentation, we use $\alpha_{n,k}^{1/2}\mathbf{h}_{n,k}$ to denote the $M$-dimensional channel vector from the BS to the $k$th MU in the $n$th cluster, where $\alpha_{n,k}$ is the large-scale channel fading, and $\mathbf{h}_{n,k}$ is the small-scale channel fading following zero mean complex Gaussian distribution with unit variance. It is assumed that $\alpha_{n,k}$ remains constant for a relatively long period, while $\mathbf{h}_{n,k}$ keeps unchanged in a time slot but varies independently over time slots.

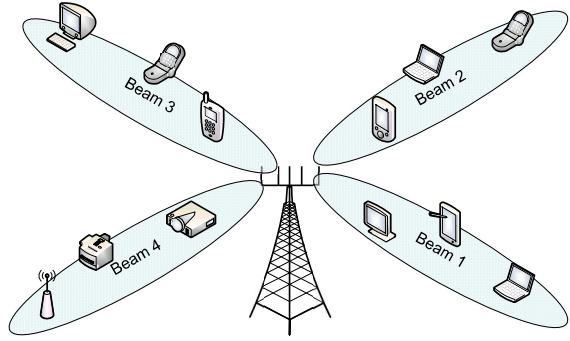

Fig. 1. A multiuser NOMA communication system with 4 clusters.

---

[1] The spatial direction of users can be found via various methods/technologies such as GPS or user location tracking algorithms.

## B. CSI Acquisition

For the TDD mode, the BS obtains the downlink CSI through uplink channel estimation. Specifically, at the beginning of each time slot, the MUs simultaneously send pilot sequences of $\tau$ symbols to the BS, and the received pilot at the BS can be expressed as

$$\mathbf{Y}_P = \sum_{n=1}^{N}\sum_{k=1}^{K} \sqrt{\tau P_{n,k}^P \alpha_{n,k}} \mathbf{h}_{n,k} \mathbf{\Phi}_{n,k} + \mathbf{N}_P, \tag{1}$$

where $P_{n,k}^P$ is the transmit power for pilot sequence of the $k$th MU in the $n$th cluster, $\mathbf{N}_P$ is an additive white Gaussian noise (AWGN) matrix with i.i.d. zero mean and unit variance complex Gaussian distributed entries. $\mathbf{\Phi}_{n,k} \in \mathbb{C}^{1\times\tau}$ is the pilot sequence sent from the $k$th MU in the $n$th cluster. It is required that $\tau > NK$, such that the pairwise orthogonality that $\mathbf{\Phi}_{n,k}\mathbf{\Phi}_{i,j}^H = 0$ and $\mathbf{\Phi}_{n,k}\mathbf{\Phi}_{n,k}^H = 1$, $\forall (n,k) \neq (i,j)$, can be guaranteed. By making use of the pairwise orthogonality, the received pilot can be transformed as

$$\mathbf{Y}_P \mathbf{\Phi}_{n,k}^H = \sqrt{\tau P_{n,k}^P \alpha_{n,k}} \mathbf{h}_{n,k} + \mathbf{N}_P \mathbf{\Phi}_{n,k}^H. \tag{2}$$

Then, by using minimum mean squared error (MMSE) estimation, the relation between the actual channel gain $\mathbf{h}_{n,k}$ and the estimated channel gain $\hat{\mathbf{h}}_{n,k}$ can be expressed as

$$\mathbf{h}_{n,k} = \sqrt{\rho_{n,k}} \hat{\mathbf{h}}_{n,k} + \sqrt{1-\rho_{n,k}} \mathbf{e}_{n,k}, \tag{3}$$

where $\mathbf{e}_{n,k}$ is the channel estimation error vector with i.i.d. zero mean and unit variance complex Gaussian distributed entries, and is independent of $\hat{\mathbf{h}}_{n,k}$. Variable $\rho_{n,k} = \frac{\tau P_{n,k}^P \alpha_{n,k}}{1+\tau P_{n,k}^P \alpha_{n,k}} = 1 - \frac{1}{1+\tau P_{n,k}^P \alpha_{n,k}}$ is the correlation coefficient between $\mathbf{h}_{n,k}$ and $\hat{\mathbf{h}}_{n,k}$. A large $\rho_{n,k}$ means a high accuracy for channel estimation. Thus, it is possible to improve the CSI accuracy by increasing the transmit power $P_{n,k}^P$ or the length $\tau$ of pilot sequence.

For the FDD mode, the CSI is usually conveyed from the MUs to the BS through a feedback link. Since the feedback link is rate-constrained, CSI at the MUs should first be quantized. Specifically, the $k$th MU in the $n$th cluster chooses an optimal codeword from a predetermined quantization codebook $\mathcal{B}_{n,k} = \{\tilde{\mathbf{h}}_{n,k}^{(1)}, \ldots, \tilde{\mathbf{h}}_{n,k}^{(2^{B_{n,k}})}\}$ of size $2^{B_{n,k}}$, where $\tilde{\mathbf{h}}_{n,k}^{(j)}$ is the $j$th codeword of unit norm and $B_{n,k}$ is the number of feedback bits. Mathematically, the codeword selection criterion is given by

$$j^\star = \arg \max_{1 \leq j \leq 2^{B_{n,k}}} \left| \mathbf{h}_{n,k}^H \tilde{\mathbf{h}}_{n,k}^{(j)} \right|^2. \tag{4}$$

Then, the MU conveys the index $j^\star$ to the BS with $B_{n,k}$ feedback bits, and the BS recoveries the quantized CSI $\tilde{\mathbf{h}}_{n,k}^{(j^\star)}$ from the same codebook. In other words, the BS only gets the phase information by using the feedback scheme based on a quantization codebook. However, as shown in below, the phase information is sufficient for the design of spatial beamforming. Similarly, the relation between the real CSI and the obtained CSI in FDD mode can be approximated as [39]

$$\tilde{\mathbf{h}}_{n,k} = \sqrt{\varrho_{n,k}} \tilde{\mathbf{h}}_{n,k}^\star + \sqrt{1-\varrho_{n,k}} \tilde{\mathbf{e}}_{n,k}, \tag{5}$$

where $\tilde{\mathbf{h}}_{n,k} = \frac{\mathbf{h}_{n,k}}{\|\mathbf{h}_{n,k}\|}$ is the phase of the channel $\mathbf{h}_{n,k}$, $\tilde{\mathbf{h}}_{n,k}^\star$ is the quantized phase information, $\tilde{\mathbf{e}}_{n,k}$ is the quantization error vector with uniform distribution, and $\varrho_{n,k} = 1 - 2^{-\frac{B_{n,k}}{M-1}}$ is the associated correlation coefficient or CSI accuracy. Thus, it is possible to improve the CSI accuracy by increasing the size of quantization codebook for a given number of antennas $M$ at the BS.

## C. Superposition Coding and Transmit Beamforming

Based on the available CSI, the BS constructs one transmit beam for each cluster, so as to mitigate or even completely cancel the inter-cluster interference. To strike balance between system performance and implementation complexity, we adopt zero-force beamforming (ZFBF) at the BS. We take the deign of beam $\mathbf{w}_i$ for the $i$th cluster as an example. First, we construct a complementary matrix $\bar{\mathbf{H}}_i$[2] as:

$$\bar{\mathbf{H}}_i = [\hat{\mathbf{h}}_{1,1}, \cdots, \hat{\mathbf{h}}_{1,K}, \cdots, \hat{\mathbf{h}}_{i-1,1}, \cdots, \hat{\mathbf{h}}_{i-1,K}, \hat{\mathbf{h}}_{i+1,1}, \cdots, \hat{\mathbf{h}}_{N,K}]^H. \tag{6}$$

Then, we perform singular value decomposition (SVD) on $\bar{\mathbf{H}}_i$ and obtain its right singular vectors $\mathbf{u}_{i,j}, j = 1, \cdots, N_u$, with respect to the zero singular values, where $N_u$ is the number of zero singular values. Finally, we can design the beam as $\mathbf{w}_i = \sum_{j=1}^{N_u} \theta_{i,j} \mathbf{u}_{i,j}$, where $\theta_{i,j} > 0$ is a weight such that $\sum_{j=1}^{N_u} \theta_{i,j} = 1$. Thus, the received signal at the $k$th MU in the $n$th cluster is given by

$$
\begin{aligned}
y_{n,k} &= \sqrt{\alpha_{n,k}} \mathbf{h}_{n,k}^H \sum_{i=1}^{N} \mathbf{w}_i s_i + n_{n,k} \\
&= \sqrt{\alpha_{n,k}} \mathbf{h}_{n,k}^H \mathbf{w}_n s_n + \sqrt{\alpha_{n,k}(1-\rho_{n,k})} \mathbf{e}_{n,k}^H \sum_{i=1,i\neq n}^{N} \mathbf{w}_i s_i + n_{n,k},
\end{aligned}
\tag{7}
$$

where $s_i = \sum_{j=1}^{K} \sqrt{P_{i,j}^S} s_{i,j}$ is the superposition coded signal with $P_{i,j}^S$ and $s_{i,j}$ being transmit power and transmit signal for the $j$th MU in the $i$th cluster, and $n_{n,k}$ is the AWGN with unit variance. In general, $P_{i,j}^S$ should be carefully allocated to distinguish the MUs in the power domain, which we will discuss in detail below. Note that Eq. (7) holds true due to the fact that $\mathbf{h}_{n,k}^H \mathbf{w}_i = \sqrt{\rho_{n,k}} \hat{\mathbf{h}}_{n,k}^H \mathbf{w}_i + \sqrt{1-\rho_{n,k}} \mathbf{e}_{n,k}^H \mathbf{w}_i = \sqrt{1-\rho_{n,k}} \mathbf{e}_{n,k}^H \mathbf{w}_i$ for ZFBF in TDD mode[3]. With perfect CSI at the BS, i.e., $\rho_{n,k} = 1$, the inter-cluster interference can be completely cancelled.

---

[2]In FDD mode, the complementary matrix is given by $\bar{\mathbf{H}}_i = [\tilde{\mathbf{h}}_{1,1}^\star, \cdots, \tilde{\mathbf{h}}_{1,K}^\star, \cdots, \tilde{\mathbf{h}}_{i-1,1}^\star, \cdots, \tilde{\mathbf{h}}_{i-1,K}^\star, \tilde{\mathbf{h}}_{i+1,1}^\star, \cdots, \tilde{\mathbf{h}}_{N,K}^\star]^H$.

### D. Successive Interference Cancellation

Although ZFBF at the BS can mitigate partial inter-cluster interference from the other clusters, there still exists intra-cluster interference from the same cluster. In order to improve the received signal quality, the MU conducts SIC according to the principle of NOMA. Without loss of generality, we assume that the effective channel gains in the $i$th cluster have the following order:

$$
|\sqrt{\alpha_{i,1}} \mathbf{h}_{i,1}^H \mathbf{w}_i|^2 \geq \cdots \geq |\sqrt{\alpha_{i,K}} \mathbf{h}_{i,K}^H \mathbf{w}_i|^2. \tag{8}
$$

It is reasonably assumed that the BS may know MUs' effective gains through the channel quality indicator (CQI) messages, and then determines the user order in (8). Thus, in the $i$th cluster, the $j$th MU can always successively decode the $l$th MU's signal, $\forall l > j$, if the $l$th MU can decode its own signal. As a result, the $j$th MU can subtract the interference from the $l$th MU in the received signal before decoding its own signal. After SIC, the signal-to-interference-plus-noise

---

[3] In FDD mode, we have $\mathbf{h}_{n,k}^H \mathbf{w}_i = \sqrt{\varrho_{n,k}} \|\mathbf{h}_{n,k}\| (\tilde{\mathbf{h}}_{n,k}^\star)^H \mathbf{w}_i + \sqrt{1-\varrho_{n,k}} \|\mathbf{h}_{n,k}\| \tilde{\mathbf{e}}_{n,k}^H \mathbf{w}_i = \sqrt{1-\varrho_{n,k}} \|\mathbf{h}_{n,k}\| \tilde{\mathbf{e}}_{n,k}^H \mathbf{w}_i \stackrel{d}{=} \sqrt{1-\varrho_{n,k}} \mathbf{e}_{n,k}^H \mathbf{w}_i$, where $\stackrel{d}{=}$ denotes the equality in distribution. If $\varrho_{n,k} = \rho_{n,k}$, Eq. (7) also holds true in FDD mode. In the sequel, without loss of generality, we no longer distinguish between TDD and FDD.

ratio (SINR) at the $k$th MU in the $n$th cluster is given by

$$\gamma_{n,k} = \frac{\alpha_{n,k}|\mathbf{h}_{n,k}^H\mathbf{w}_n|^2 P_{n,k}^S}{\underbrace{\alpha_{n,k}|\mathbf{h}_{n,k}^H\mathbf{w}_n|^2 \sum_{j=1}^{k-1} P_{n,j}^S}_{\text{Intra-cluster interference}} + \underbrace{\alpha_{n,k}(1-\rho_{n,k})\sum_{i=1,i\neq n}^{N}|\mathbf{e}_{n,k}^H\mathbf{w}_i|^2 \sum_{l=1}^{K} P_{i,l}^S}_{\text{Inter-cluster interference}} + \underbrace{1}_{\text{AWGN}}}, \quad (9)$$

where the first term in the denominator of (9) is the residual intra-cluster interference after SIC at the MU, the second one is the residual inter-cluster interference after ZFBF at the BS, and the third one is the AWGN. For the 1st MU in each cluster, there is no intra-cluster interference, since it can completely eliminate the intra-cluster interference. Note that in this paper, we assume that perfect SIC can be performed at the MUs. In practical NOMA systems, SIC might be imperfect due to a limited computational capability at the MUs. Thus, there exists residual intra-cluster interference from the weaker MUs even after SIC [37]. However, the study of the impact of imperfect SIC on the system performance is beyond the scope of this paper and we would like to investigate it in the future work. Moreover, the transmit power has a significant impact on the SIC and the performance of NOMA [38]. Thus, we will quantitatively analyze the impact of transmit power and then aim to optimize the transmit power for improving the performance in the following sections.

## III. Performance Analysis and Optimization

In this section, we concentrate on performance analysis and optimization of multi-antenna NOMA downlink with imperfect CSI. Specifically, we first derive closed-form expressions for the average transmission rates of the 1st MU and the other MUs, and then propose separate and joint optimization schemes of transmit power, feedback bits, and transmit mode, so as to maximize the average sum rate of the system.

### A. Average Transmission Rate

We start by analyzing the average transmission rate of the $k$th MU in the $n$th cluster. First, we consider the case $k > 1$. According to the definition, the corresponding average transmission rate can be computed as

$$R_{n,k} = \mathrm{E}\left[\log_2\left(1 + \gamma_{n,k}\right)\right]$$

$$= \mathrm{E}\left[\log_2\left(\frac{\alpha_{n,k}|\mathbf{h}_{n,k}^H\mathbf{w}_n|^2 \sum_{j=1}^{k} P_{n,j}^S + \alpha_{n,k}(1-\rho_{n,k})\sum_{i=1,i\neq n}^{N}|\mathbf{e}_{n,k}^H\mathbf{w}_i|^2\sum_{l=1}^{K}P_{i,l}^S+1}{\alpha_{n,k}|\mathbf{h}_{n,k}^H\mathbf{w}_n|^2 \sum_{j=1}^{k-1} P_{n,j}^S + \alpha_{n,k}(1-\rho_{n,k})\sum_{i=1,i\neq n}^{N}|\mathbf{e}_{n,k}^H\mathbf{w}_i|^2\sum_{l=1}^{K}P_{i,l}^S+1}\right)\right]$$

$$= \mathrm{E}\left[\log_2\left(\alpha_{n,k}|\mathbf{h}_{n,k}^H\mathbf{w}_n|^2 \sum_{j=1}^{k} P_{n,j}^S + \alpha_{n,k}(1-\rho_{n,k})\sum_{i=1,i\neq n}^{N}|\mathbf{e}_{n,k}^H\mathbf{w}_i|^2\sum_{l=1}^{K}P_{i,l}^S+1\right)\right]$$

$$-\mathrm{E}\left[\log_2\left(\alpha_{n,k}|\mathbf{h}_{n,k}^H\mathbf{w}_n|^2 \sum_{j=1}^{k-1} P_{n,j}^S + \alpha_{n,k}(1-\rho_{n,k})\sum_{i=1,i\neq n}^{N}|\mathbf{e}_{n,k}^H\mathbf{w}_i|^2\sum_{l=1}^{K}P_{i,l}^S+1\right)\right]. \quad (10)$$

Note that the average transmission rate in (10) can be expressed as the difference of two terms, which have a similar form. Hence, we concentrate on the derivation of the first term. For notational convenience, we use $W$ to denote the term $\alpha_{n,k}|\mathbf{h}_{n,k}^H\mathbf{w}_n|^2\sum_{j=1}^{k}P_{n,j}^S + \alpha_{n,k}(1-\rho_{n,k})\sum_{i=1,i\neq n}^{N}|\mathbf{e}_{n,k}^H\mathbf{w}_i|^2\sum_{l=1}^{K}P_{i,l}^S$. To compute the first expectation, the key is to obtain the probability density function (pdf) of $W$. Checking the first random variable $|\mathbf{h}_{n,k}^H\mathbf{w}_n|^2$ in $W$, since $\mathbf{w}_n$ of unit norm is designed independent of $\mathbf{h}_{n,k}$, $|\mathbf{h}_{n,k}^H\mathbf{w}_n|^2$ is $\chi^2$ distributed with 2 degrees of freedom [40]. Similarly, $|\mathbf{e}_{n,k}^H\mathbf{w}_i|^2$ also has the distribution $\chi^2(2)$. Therefore, $W$ can be considered as a weighted sum of $N$ random variables with $\chi^2(2)$ distribution. According to [41], $W$ is a nested finite weighted sum of $N$ Erlang pdfs, whose pdf is given by

$$f_W(x) = \sum_{i=1}^{N}\Xi_N\left(i,\{\eta_{n,k}^q\}_{q=1}^N\right)g(x,\eta_{n,k}^i), \quad (11)$$

where

$$\eta_{n,k}^q = \begin{cases}\alpha_{n,k}\sum_{j=1}^{k}P_{q,j}^S & \text{if } q=n\\ \alpha_{n,k}(1-\rho_{n,k})\sum_{l=1}^{K}P_{q,l}^S & \text{if } q\neq n\end{cases},$$

$$g(x,\eta_{n,k}^i) = \frac{1}{\eta_{n,k}^i}\exp\left(-\frac{x}{\eta_{n,k}^i}\right),$$

$$\Xi_N\left(i,\{\eta_{n,k}^q\}_{q=1}^N\right) = \frac{(-1)^{N-1}\eta_{n,k}^i}{\prod_{l=1}^{N}\eta_{n,k}^l}\prod_{s=1}^{N-1}\left(\frac{1}{\eta_{n,k}^i} - \frac{1}{\eta_{n,k}^{s+\mathbf{U}(s-i)}}\right)^{-1},$$

and $\mathbf{U}(x)$ is the well-known unit step function defined as $\mathbf{U}(x \geq 0) = 1$ and zero otherwise. It is worth pointing out that the weights $\Xi_N$ are constant for given $\{\eta_{n,k}^q\}_{q=1}^N$. Hence, the first expectation in (10) can be computed as

$$\begin{aligned}
\mathrm{E}[\log_2(1+W)] &= \int_0^\infty \log_2(1+x) f_W(x) dx \\
&= \sum_{i=1}^N \Xi_N\left(i, \{\eta_{n,k}^q\}_{q=1}^N\right) \int_0^\infty \log_2(1+x) \frac{1}{\eta_{n,k}^i} \exp\left(-\frac{x}{\eta_{n,k}^i}\right) dx \\
&= -\frac{1}{\ln(2)} \sum_{i=1}^N \Xi_N\left(i, \{\eta_{n,k}^q\}_{q=1}^N\right) \exp\left(\frac{1}{\eta_{n,k}^i}\right) \mathrm{E_i}\left(-\frac{1}{\eta_{n,k}^i}\right),
\end{aligned} \quad (12)$$

where $\mathrm{E_i}(x) = \int_{-\infty}^x \frac{\exp(t)}{t} dt$ is the exponential integral function. Eq. (12) follows from [42, Eq. (4.3372)]. Similarly, we use $V$ to denote $\alpha_{n,k} |\mathbf{h}_{n,k}^H \mathbf{w}_n|^2 \sum_{j=1}^{k-1} P_{n,j}^S + \alpha_{n,k}(1-\rho_{n,k}) \sum_{i=1, i \neq n}^N |\mathbf{e}_{n,k}^H \mathbf{w}_i|^2 \sum_{t=1}^K P_{i,t}^S$ in the second term of (10). Thus, the second expectation term can be computed as

$$\mathrm{E}[\log_2(1+V)] = -\frac{1}{\ln(2)} \sum_{i=1}^N \Xi_N\left(i, \{\beta_{n,k}^v\}_{v=1}^N\right) \exp\left(\frac{1}{\beta_{n,k}^i}\right) \mathrm{E_i}\left(-\frac{1}{\beta_{n,k}^i}\right), \quad (13)$$

where

$$\beta_{n,k}^v = \begin{cases} \alpha_{n,k} \sum_{j=1}^{k-1} P_{v,j}^S & \text{if } v = n \\ \alpha_{n,k}(1-\rho_{n,k}) \sum_{l=1}^K P_{v,l}^S & \text{if } v \neq n \end{cases}.$$

Hence, we can obtain the average transmission rate for the $k$th MU in the $n$th cluster as follows

$$\begin{aligned}
R_{n,k} &= \frac{1}{\ln(2)} \sum_{i=1}^N \Xi_N\left(i, \{\beta_{n,k}^v\}_{v=1}^N\right) \exp\left(\frac{1}{\beta_{n,k}^i}\right) \mathrm{E_i}\left(-\frac{1}{\beta_{n,k}^i}\right) \\
&\quad - \frac{1}{\ln(2)} \sum_{i=1}^N \Xi_N\left(i, \{\eta_{n,k}^q\}_{q=1}^N\right) \exp\left(\frac{1}{\eta_{n,k}^i}\right) \mathrm{E_i}\left(-\frac{1}{\eta_{n,k}^i}\right).
\end{aligned} \quad (14)$$

Then, we consider the case $k = 1$. Since the first MU can decode all the other MUs' signals in the same cluster, there is no intra-cluster interference. In this case, the corresponding average transmission rate reduces to

$$R_{n,1} = \frac{1}{\ln(2)} \sum_{i=1}^{N-1} \Xi_{N-1}\left(i, \{\beta_{n,1}^v\}_{v=1}^{N-1}\right) \exp\left(\frac{1}{\beta_{n,1}^i}\right) \mathrm{E_i}\left(-\frac{1}{\beta_{n,1}^i}\right)$$

$$-\frac{1}{\ln(2)} \sum_{i=1}^{N} \Xi_N \left(i, \{\eta_{n,1}^q\}_{q=1}^N\right) \exp\left(\frac{1}{\eta_{n,1}^i}\right) \mathrm{E_i}\left(-\frac{1}{\eta_{n,1}^i}\right), \quad (15)$$

where

$$\eta_{n,1}^q = \begin{cases} \alpha_{n,1} P_{q,1}^S & \text{if } q = n \\ \alpha_{n,1}(1-\rho_{n,1}) \sum_{l=1}^{K} P_{q,l}^S & \text{if } q \neq n \end{cases},$$

and

$$\beta_{n,1}^v = \begin{cases} \alpha_{n,1}(1-\rho_{n,1}) \sum_{l=1}^{K} P_{v,l}^S & \text{if } v < n \\ \alpha_{n,1}(1-\rho_{n,1}) \sum_{l=1}^{K} P_{v+1,l}^S & \text{if } v \geq n \end{cases}.$$

Combing (14) and (15), it is easy to evaluate the performance of a multiple-antenna NOMA downlink with arbitrary system parameters and channel conditions. In particular, it is possible to reveal the impact of system parameters, i.e., transmit power, CSI accuracy, and transmission mode.

## B. Power Allocation

From (14) and (15), it is easy to observe that with imperfect CSI, transmit power has a great impact on average transmission rates. On one hand, increasing the transmit power can enhance the desired signal strength. On the other hand, it also increases the interference. Thus, it is desired to distribute the transmit power according to channel conditions.

To maximize the sum rate of the considered multiple-antenna NOMA system subject to a total power constraint, we have the following optimization problem:

$$J_1 : \max_{P_{n,k}^S} \sum_{n=1}^{N} \sum_{k=1}^{K} R_{n,k} \quad (16)$$

$$\text{s.t.} \quad \text{C1}: \sum_{n=1}^{N} \sum_{k=1}^{K} P_{n,k}^S \leq P_{tot}^S$$

$$\text{C2}: P_{n,k}^S > 0,$$

where $P_{tot}^S$ is the maximum total transmit power budget. It is worth pointing out that in certain scenarios, user fairness might be of particular importance. To guarantee user fairness, one can replace the objective function of $J_1$ with the maximization of a weighted sum rate, where the weights can directly affect the power allocation and thus the MUs' rates. Unfortunately, $J_1$ is not a convex problem due to the complicated expression for the objective function. Thus, it is difficult to directly provide a closed-form solution for the optimal transmit power. As a compromise solution, we propose an effective power allocation scheme based on the following important observation of the multiple-antenna NOMA downlink system:

*Lemma 1*: The inter-cluster interference is dependent of power allocation between the clusters, while the intra-cluster interference is determined by power allocation among the MUs in the same cluster.

*Proof:* A close observation of the inter-cluster interference $\alpha_{n,k}(1-\rho_{n,k})\sum_{i=1,i\neq n}^{N}|\mathbf{e}_{n,k}^H\mathbf{w}_i|^2 \sum_{l=1}^{K} P_{i,l}^S$ in (9) indicates that $\sum_{l=1}^{K} P_{i,l}^S$ is the total transmit power for the $i$th cluster, which suggests that inter-cluster power allocation does not affect the inter-cluster interference. ∎

Inspired by Lemma 1, the power allocation scheme can be divided into two steps. In the first step, the BS distributes the total power among the $N$ clusters. In the second step, each cluster individually carries out power allocation subject to the power constraint determined by the first step. In the following, we give the details of the two-step power allocation scheme. First, we design the power allocation between the clusters from the perspective of minimizing inter-cluster interference. For the $i$th cluster, the average aggregate interference to the other clusters is given by

$$\begin{aligned} I_i &= \mathrm{E}\left[\sum_{n=1,n\neq i}^{N}\sum_{k=1}^{K}\alpha_{n,k}(1-\rho_{n,k})|\mathbf{e}_{n,k}^H\mathbf{w}_i|^2 \sum_{l=1}^{K} P_{i,l}^S\right] \\ &= \left(\sum_{n=1,n\neq i}^{N}\sum_{k=1}^{K}\alpha_{n,k}(1-\rho_{n,k})\right) P_i^S, \end{aligned} \quad (17)$$

where $P_i^S = \sum_{l=1}^{K} P_{i,l}^S$ is the total transmit power of the $i$th cluster. Eq. (17) follows the fact that $\mathrm{E}[|\mathbf{e}_{n,k}^H\mathbf{w}_i|^2] = 1$. Intuitively, a large interference coefficient $\sum_{n=1,n\neq i}^{N}\sum_{k=1}^{K}\alpha_{n,k}(1-\rho_{n,k})$ means a more severe inter-cluster interference caused by the $i$th cluster. In order to mitigate the inter-cluster interference for improving the average sum rate, we propose to distribute the power proportionally to the reciprocal of interference coefficient. Specifically, the transmit power for

the $i$th cluster can be computed as

$$P_i^S = \frac{\left(\sum_{n=1,n\neq i}^{N}\sum_{k=1}^{K}\alpha_{n,k}(1-\rho_{n,k})\right)^{-1}}{\sum_{l=1}^{N}\left(\sum_{n=1,n\neq l}^{N}\sum_{k=1}^{K}\alpha_{n,k}(1-\rho_{n,k})\right)^{-1}}P_{tol}^S. \quad (18)$$

Then, we allocate the power in the cluster for further increasing the average sum rate. According to the nature of NOMA techniques, the first MU not only has the strongest effective channel gain for the desired signal, but also generates a weak interference to the other MUs. On the contrary, the $K$th MU has the weakest effective channel gain for the desired signal, and also produces a strong interference to the other MUs. Thus, from the perspective of maximizing the sum of average rate, it is better to allocate the power based on the following criterion:

$$P_{n,1}^S \geq \cdots \geq P_{n,k}^S \geq \cdots \geq P_{n,K}^S. \quad (19)$$

On the other hand, in order to facilitate SIC, the NOMA in general requires the transmit powers in a cluster to follow a criterion below [28]:

$$P_{n,1}^S \leq \cdots \leq P_{n,k}^S \leq \cdots \leq P_{n,K}^S. \quad (20)$$

Under this condition, the MU performs SIC according to the descending order of the user index, namely the ascending order of the effective channel gain. Specifically, the $k$th MU cancels the interference from the $K$th to the $(k+1)$th MU in sequence. Thus, the SINR for decoding each interference signal is the highest, which facilitates SIC at MUs [38].

To simultaneously fulfill the above two criterions, we propose to equally distribute the powers within a cluster, namely

$$P_{n,k}^S = P_n^S/K. \quad (21)$$

Substituting (18) into (21), the transmit power for the $k$th MU in the $n$th cluster can be computed

as

$$P_{n,k}^S = \frac{\left(\sum_{i=1,i\neq n}^{N}\sum_{j=1}^{K}\alpha_{i,j}(1-\rho_{i,j})\right)^{-1}}{K\left(\sum_{l=1}^{N}\left(\sum_{i=1,i\neq l}^{N}\sum_{j=1}^{K}\alpha_{i,j}(1-\rho_{i,j})\right)^{-1}\right)}P_{tol}^S. \quad (22)$$

Thus, we can distribute the transmit power based on (22) for given channel statistical information and the CSI accuracy, which has a quite low computational complexity.

*Remarks*: We note that path loss coefficient $\alpha_{n,k}, \forall n,k$, remain constant for a relatively long time, and it is easy to obtain at the BS via long-term measurement. Hence, the proposed power allocation scheme incurs a low system overhead, and can be implemented with low complexity.

## C. Feedback Distribution

For the FDD mode, the accuracy of quantized CSI relies on the size of codebook $2^{B_{n,k}}$, where $B_{n,k}$ is the number of feedback bits from the $k$th MU in the $n$th cluster. As observed in (14) and (15), it is possible to decrease the interference by increasing feedback bits. However, due to the rate constraint on the feedback link, the total number of feedback bits is limited. Therefore, it is of great importance to optimize the feedback bits among the MUs for performance enhancement.

According to the received SNR in (9), the CSI accuracy only affects the inter-cluster interference. Thus, it makes sense to optimize the feedback bits to minimizing the average sum of inter-cluster interference given by

$$\begin{aligned} I_{\text{inter}} &= \mathrm{E}\left[\sum_{n=1}^{N}\sum_{k=1}^{K}\alpha_{n,k}(1-\varrho_{n,k})\sum_{i=1,i\neq n}^{N}|\mathbf{e}_{n,k}^H\mathbf{w}_i|^2\sum_{l=1}^{K}P_{i,l}^S\right] \\ &= \sum_{n=1}^{N}\sum_{k=1}^{K}\alpha_{n,k}\sum_{i=1,i\neq n}^{N}P_i^S 2^{-\frac{B_{n,k}}{M-1}}. \end{aligned} \quad (23)$$

Hence, the optimization problem for feedback bits distribution can be expressed as

$$J_2 : \min_{B_{n,k}} \sum_{n=1}^{N} \sum_{k=1}^{K} \alpha_{n,k} \sum_{i=1,i\neq n}^{N} P_i^S 2^{-\frac{B_{n,k}}{M-1}} \tag{24}$$

$$\text{s.t.} \quad C3 : \sum_{n=1}^{N} \sum_{k=1}^{K} B_{n,k} \leq B_{\text{tot}},$$

$$C4 : B_{n,k} \geq 0,$$

where $B_{\text{tot}}$ is an upper bound on the total number of feedback bits. $J_2$ is an integer programming problem, hence is difficult to solve. To tackle this challenge, we relax the integer constraint on $B_{n,k}$. In this case, according to the fact that

$$\sum_{n=1}^{N} \sum_{k=1}^{K} \alpha_{n,k} \sum_{i=1,i\neq n}^{N} P_i^S 2^{-\frac{B_{n,k}}{M-1}} \geq NK \left( \prod_{n=1}^{N} \prod_{k=1}^{K} \alpha_{n,k} \sum_{i=1,i\neq n}^{N} P_i^S 2^{-\frac{B_{n,k}}{M-1}} \right)^{\frac{1}{NK}}$$

$$= NK \left( 2^{-\frac{\sum_{n=1}^{N}\sum_{k=1}^{K} B_{n,k}}{M-1}} \right)^{\frac{1}{NK}} \left( \prod_{n=1}^{N} \prod_{k=1}^{K} \alpha_{n,k} \sum_{i=1,i\neq n}^{N} P_i^S \right)^{\frac{1}{NK}}$$

$$= NK \left( 2^{-\frac{B_{tot}}{M-1}} \right)^{\frac{1}{NK}} \left( \prod_{n=1}^{N} \prod_{k=1}^{K} \alpha_{n,k} \sum_{i=1,i\neq n}^{N} P_i^S \right)^{\frac{1}{NK}}, \tag{25}$$

where the equality holds true only when $\alpha_{n,k} \sum_{i=1,i\neq n}^{N} P_i^S 2^{-\frac{B_{n,k}}{M-1}}, \forall n, k$ are equal. In other words, the objective function in (24) can be minimized while satisfying the following condition:

$$\alpha_{n,k} \sum_{i=1,i\neq n}^{N} P_i^S 2^{-\frac{B_{n,k}}{M-1}} = \left( 2^{-\frac{B_{tot}}{M-1}} \right)^{\frac{1}{NK}} \left( \prod_{n=1}^{N} \prod_{k=1}^{K} \alpha_{n,k} \sum_{i=1,i\neq n}^{N} P_i^S \right)^{\frac{1}{NK}}. \tag{26}$$

Hence, based on the relaxed optimization problem, the optimal number of feedback bits for the $k$th MU in the $n$th cluster is given by

$$B_{n,k} = \frac{B_{\text{tot}}}{NK} - \frac{1}{NK} \sum_{i=1}^{N} \sum_{j=1}^{K} \log_2 \left( \alpha_{i,j} \sum_{l=1,l\neq i}^{N} P_l^S \right) + \log_2 \left( \alpha_{n,k} \sum_{l=1,l\neq n}^{N} P_l^S \right). \tag{27}$$

Given channel statistical information and transmit power allocation, it is easy to determine the feedback distribution according to (27). Note that there exists an integer constraint on the number of feedback bits in practice, so we should utilize the maximum integer that is not larger than

$B_{n,k}$ in (27), i.e., $\lfloor B_{n,k} \rfloor, \forall n, k$.

*Remarks*: The number of feedback bits distributed to the $k$th MU in the $n$th cluster is determined by the average inter-cluster interference generated by the $k$th MU in the $n$th cluster with respect to the average inter-cluster interference of each MU. In other words, if one MU generates more inter-cluster interference, it would be allocated with more feedback bits, so as to facilitate a more accurate ZFBF to minimize the total interference.

### D. Mode Selection

As discussed above, the performance of the multiple-antenna NOMA system is limited by both inter-cluster and intra-cluster interference. Although ZFBF at the BS and SIC at the MUs are jointly applied, there still exists residual interference. Intuitively, the strength of the residual interference mainly relies on the number of clusters $N$ and the number of MUs in each cluster $K$. For instance, increasing the number of MUs in each cluster might reduce the inter-cluster interference, but also results in an increase in intra-cluster interference. Thus, it is desired to dynamically adjust the transmission mode, including the number of clusters and the number of MUs in each cluster, according to channel conditions and system parameters. For dynamic mode selection, we have the following lemma:

*Lemma 2*: If the BS has no CSI about the downlink, it is optimal to set $N = 1$. On the other hand, if the BS has perfect CSI about the downlink, $K = 1$ is the best choice.

*Proof:* First, if there is no CSI, namely $\rho_{n,k} = 0, \forall n, k$, ZFBF cannot be utilized to mitigate the inter-cluster interference. If all the MUs belong to one cluster, interference can be mitigated as much as possible by SIC. In the case of perfect CSI at the BS, ZFBF can completely the interference. Thus, it is optimal to arrange one MU in one cluster. ∎

In above, we consider two extreme scenarios of no and perfect CSI at the BS, respectively. In practice, the BS has partial CSI through channel estimation or quantization feedback. Thus, we propose to dynamically choose the transmission mode for maximizing the sum of average

transmission rate, which is equivalent to an optimization problem below:

$$J_3 \; : \; \max_{N,K} \sum_{n=1}^{N} \sum_{k=1}^{K} R_{n,k} \quad (28)$$

$$\text{s.t.} \quad \text{C5}: \quad NK = N_u,$$

$$\text{C6}: \quad N > 0,$$

$$\text{C7}: \quad K > 0,$$

where $N_u$ is the number of MUs in the multiple-antenna NOMA system. $J_3$ is also an integer programming problem, so it is difficult to obtain the closed-form solution. Under this condition, it is feasible to get the optimal solution by numerical search and the search complexity is $O(N^K)$. In order to control the complexity of SIC, the number of MUs in one cluster is usually small, e.g., $K = 2$. Therefore, the complexity of numerical search is acceptable.

*E. Joint Optimization Scheme*

In fact, transmit power, feedback bits and transmission mode are coupled, and determine the performance together. Therefore, it is better to jointly optimize these variables, so as to further improve the performance of the multiple-antenna NOMA systems. For example, given a transmission mode, it is easy to first allocate transmit power according to (22), and then distribute feedback bits according to (27). Finally, we can select an optimal transmission mode with the largest sum rate. The complexity of the joint optimization is mainly determined by the mode selection. As mentioned above, if the number of MUs in one cluster is small, the complex of mode selection is acceptable.

## IV. ASYMPTOTIC ANALYSIS

In order to provide insightful guidelines for system design, we now pursue an asymptotic analysis on the average sum rate of the system. In particular, two extreme cases are studied, namely, interference limited and noise limited.

## A. Interference Limited Case

With loss of generality, we let $P_{n,k}^S = \theta_{n,k} P_{tot}^S, \forall n, k$, where $0 < \theta_{n,k} < 1$ is a power allocation factor. For instance, $\theta_{n,k}$ is equal to $\dfrac{\left(\sum\limits_{v=1,v\neq n}^{N}\sum\limits_{j=1}^{K}\alpha_{v,j}(1-\rho_{v,j})\right)^{-1}}{K\left(\sum\limits_{l=1}^{N}\left(\sum\limits_{v=1,v\neq l}^{N}\sum\limits_{j=1}^{K}\alpha_{v,j}(1-\rho_{v,j})\right)^{-1}\right)}$ in the proposed power allocation scheme in Section III.B. If the total power $P_{tot}^S$ is large enough, the noise term of SINR in (9) is negligible. In this case, with the help of [42, Eq. (4.3311)], the average transmission rate of the $k$th MU ($k > 1$) in the $n$th cluster reduces to

$$R_{n,k} = \frac{1}{\ln(2)} \sum_{i=1}^{N} \Xi_N\left(i, \{\eta_{n,k}^q\}_{q=1}^N\right) \ln(\eta_{n,k}^i) - \frac{1}{\ln(2)} \sum_{i=1}^{N} \Xi_N\left(i, \{\beta_{n,k}^v\}_{v=1}^N\right) \ln(\beta_{n,k}^i), \qquad (29)$$

where we have also used the fact that

$$\sum_{i=1}^{N} \Xi_N\left(i, \{\eta_{n,k}^q\}_{q=1}^N\right) = \sum_{i=1}^{N} \Xi_N\left(i, \{\beta_{n,k}^v\}_{v=1}^N\right) = 1. \qquad (30)$$

Similarly, the asymptotic average transmission rate of the 1st MU in the $n$th MU can be obtained as

$$R_{n,1} = \frac{1}{\ln(2)} \sum_{i=1}^{N} \Xi_N\left(i, \{\eta_{n,1}^q\}_{q=1}^N\right) \ln\left(\eta_{n,1}^i\right) - \frac{1}{\ln(2)} \sum_{i=1}^{N-1} \Xi_{N-1}\left(i, \{\beta_{n,1}^v\}_{v=1}^{N-1}\right) \ln\left(\beta_{n,1}^i\right). \qquad (31)$$

Combining (29) and (31), we have the following important result:

*Theorem 1*: In the region of high transmit power, the average transmission rate is independent of $P_{tot}^S$, and there exists a performance ceiling regardless of $P_{tot}^S$, i.e., once $P_{tot}^S$ is larger than a saturation point, the average transmission rate will not increase further even the transmit power increases.

*Proof:* According to the definitions, $\eta_{n,k}^i$ and $\beta_{n,k}^i$ can be rewritten as $\eta_{n,k}^i = \omega_{n,k}^i P_{tot}^S$ and $\beta_{n,k}^i = \psi_{n,k}^i P_{tot}^S$, where

$$\omega_{n,k}^i = \begin{cases} \alpha_{n,k} \sum\limits_{j=1}^{k} \theta_{i,j} & \text{if } i = n \\ \alpha_{n,k}(1-\rho_{n,k}) \sum\limits_{l=1}^{K} \theta_{i,l} & \text{if } i \neq n \end{cases},$$

and
$$\psi_{n,k}^i = \begin{cases} \alpha_{n,k} \sum_{j=1}^{k-1} \theta_{i,j} & \text{if } i = n \\ \alpha_{n,k}(1-\rho_{n,k}) \sum_{l=1}^{K} \theta_{i,l} & \text{if } i \neq n \end{cases},$$

respectively. Thus, $\Xi_N\left(i, \{\eta_{n,k}^q\}_{q=1}^N\right)$ and $\Xi_N\left(i, \{\beta_{n,k}^v\}_{v=1}^N\right)$ are independent of $P_{tot}^S$. Hence, $R_{n,k}$ in (29) can be transformed as

$$\begin{aligned} R_{n,k} &= \frac{1}{\ln(2)} \sum_{i=1}^{N} \Xi_N\left(i, \{\eta_{n,k}^q\}_{q=1}^N\right) \left(\ln(P_{tot}^S) + \ln(\omega_{n,k}^i)\right) \\ &\quad - \frac{1}{\ln(2)} \sum_{i=1}^{N} \Xi_N\left(i, \{\beta_{n,k}^v\}_{v=1}^N\right) \left(\ln(P_{tot}^S) + \ln(\psi_{n,k}^i)\right) \\ &= \frac{1}{\ln(2)} \sum_{i=1}^{N} \Xi_N\left(i, \{\eta_{n,k}^q\}_{q=1}^N\right) \ln(\omega_{n,k}^i) - \frac{1}{\ln(2)} \sum_{i=1}^{N} \Xi_N\left(i, \{\beta_{n,k}^v\}_{v=1}^N\right) \ln(\psi_{n,k}^i), \quad (32) \end{aligned}$$

where Eq. (32) follows the fact that $\sum_{i=1}^{N} \Xi_N\left(i, \{\eta_{n,k}^q\}_{q=1}^N\right) = \sum_{i=1}^{N} \Xi_N\left(i, \{\beta_{n,k}^v\}_{v=1}^N\right) = 1$. Similarly, we can rewrite $R_{n,1}$ in (31) as

$$R_{n,1} = \frac{1}{\ln(2)} \sum_{i=1}^{N} \Xi_N\left(i, \{\eta_{n,1}^q\}_{q=1}^N\right) \ln\left(\omega_{n,1}^i\right) - \frac{1}{\ln(2)} \sum_{i=1}^{N-1} \Xi_{N-1}\left(i, \{\beta_{n,1}^v\}_{v=1}^{N-1}\right) \ln\left(\psi_{n,1}^i\right), \quad (33)$$

where
$$\omega_{n,1}^i = \begin{cases} \alpha_{n,1} \theta_{i,1}^S & \text{if } i = n \\ \alpha_{n,1}(1-\rho_{n,1}) \sum_{l=1}^{K} \theta_{i,l}^S & \text{if } i \neq n \end{cases},$$

and
$$\psi_{n,1}^i = \begin{cases} \alpha_{n,1}(1-\rho_{n,1}) \sum_{l=1}^{K} \theta_{i,l}^S & \text{if } i < n \\ \alpha_{n,1}(1-\rho_{n,1}) \sum_{l=1}^{K} \theta_{i+1,l}^S & \text{if } i \geq n \end{cases}.$$

Note that both (32) and (33) are regardless of $P_{tot}^S$, which proves Theorem 1. ∎

Now, we investigate the relation between the performance ceiling in Theorem 1 and the CSI accuracy $\rho_{n,k}$. First, we consider $R_{n,k}$ with $k > 1$. As $\rho_{n,k}$ asymptotically approaches 1, the inter-cluster interference is negligible. Then, $R_{n,k}$ can be further reduced as

$$\begin{aligned} R_{n,k}^{\text{ideal}} &= \text{E}\left[\log_2\left(\alpha_{n,k}|\mathbf{h}_{n,k}^H\mathbf{w}_n|^2\sum_{j=1}^{k}P_{n,j}^S\right)\right] - \text{E}\left[\log_2\left(\alpha_{n,k}|\mathbf{h}_{n,k}^H\mathbf{w}_n|^2\sum_{j=1}^{k-1}P_{n,j}^S\right)\right] \\ &= \log_2\left(\frac{\sum_{j=1}^{k}\omega_{n,j}}{\sum_{j=1}^{k-1}\psi_{n,j}}\right). \end{aligned} \quad (34)$$

It is found that even with perfect CSI, the average transmission rate for the $(k>1)$th MU is still upper bounded. The bound $\log_2\left(\frac{\sum_{j=1}^{k}\omega_{n,j}}{\sum_{j=1}^{k-1}\psi_{n,j}}\right)$ is completely determined by channel conditions, and thus cannot be increased via power allocation. Differently, for the 1st MU, if the CSI at the BS is sufficiently accurate, the SINR $\gamma_{n,1}$ becomes high. As a result, the constant term 1 in the rate expression is negligible, and thus the average transmission rate can be approximated as

$$\begin{aligned} R_{n,1} &\approx \text{E}\left[\log_2\left(\frac{\alpha_{n,1}|\mathbf{h}_{n,1}^H\mathbf{w}_n|^2 P_{n,1}^S}{\alpha_{n,1}(1-\rho_{n,1})\sum_{i=1,i\neq n}^{N}|\mathbf{e}_{n,1}^H\mathbf{w}_i|^2\sum_{l=1}^{K}P_{i,l}^S}\right)\right] \\ &= \underbrace{\text{E}\left[\log_2\left(\alpha_{n,1}|\mathbf{h}_{n,1}^H\mathbf{w}_n|^2 P_{n,1}^S\right)\right]}_{\text{Ideal average rate}} - \underbrace{\text{E}\left[\log_2\left(\alpha_{n,1}(1-\rho_{n,1})\sum_{i=1,i\neq n}^{N}|\mathbf{e}_{n,1}^H\mathbf{w}_i|^2\sum_{l=1}^{K}P_{i,l}^S\right)\right]}_{\text{Rate loss due to imperfect CSI}}.(35) \end{aligned}$$

In (35), the first term is the ideal average transmission rate with perfect CSI, and the second one is rate loss caused by imperfect CSI. We first check the term of the ideal average transmission rate, which is given by

$$\begin{aligned} R_{n,1}^{\text{ideal}} &= \text{E}\left[\log_2\left(\alpha_{n,1}P_{tot}^S\theta_{n,1}|\mathbf{h}_{n,1}^H\mathbf{w}_n|^2\right)\right] \\ &= \log_2\left(\alpha_{n,1}P_{tot}^S\theta_{n,1}\right) - \frac{C}{\ln(2)}. \end{aligned} \quad (36)$$

Note that if there is perfect CSI at the BS, the average transmission rate of the 1st MU increases proportionally to $\log_2(P_{tot}^S)$ without a bound. However, as seen in (34), the $(k>1)$th MU has an upper bounded rate under the same condition, which reconfirms the claim in Lemma 2 that it is optimal to arrange one MU in each cluster in presence of perfect CSI. Then, we investigate the rate loss due to imperfect CSI, which can be expressed as

$$R_{n,1}^{\text{loss}} = \log_2\left(\alpha_{n,1}(1-\rho_{n,1})P_{tot}^S\right) - \frac{1}{\ln(2)}\sum_{i=1}^{N-1}\Xi_{N-1}\left(i,\{\mu_{n,1}^v\}_{v=1}^{N-1}\right)\left(C - \ln\left(\mu_{n,1}^i\right)\right), \quad (37)$$

where

$$\mu_{n,1}^v = \begin{cases} \sum_{l=1}^{K}\theta_{v,l} & \text{if } v < n \\ \sum_{l=1}^{K}\theta_{v+1,l} & \text{if } v \geq n \end{cases}.$$

Given a $\rho_{n,1}$, the rate loss $R_{n,1}^{loss}$ enlarges as the total transmit power $P_{tot}^S$ increases. In order to keep the same rate of increase to the ideal rate $R_{n,1}^{ideal}$, the CSI accuracy $\rho_{n,1}$ should satisfy the following theorem:

*Theorem 2*: Only when $(1-\rho_{n,1})P_{tot}^S$ is equal to a constant $\varepsilon$, the average transmission rate of the 1st MU in the $n$th cluster with imperfect CSI remains a fixed gap with respect to the ideal rate. Specifically, the transmit power for training sequence should satisfy $P_{n,1}^p = \frac{P_{tot}^S/\varepsilon - 1}{\alpha_{n,1}\tau}$ in TDD systems, while the number of feedback bits should satisfy $B_{n,1} = (M-1)\log_2(P_{tot}^S/\varepsilon)$ in FDD systems.

*Proof:* The proof is intuitively. By substituting $\rho_{n,1} = 1 - \frac{1}{1+\tau P_{n,1}^P \alpha_{n,1}}$ into $(1-\rho_{n,1})P_{tot}^S = \varepsilon$ for TDD systems and $\varrho_{n,1} = 1 - 2^{-\frac{B_{n,1}}{M-1}}$ into $(1-\varrho_{n,1})P_{tot}^S = \varepsilon$ for FDD systems, we can get $P_{n,1}^p = \frac{P_{tot}^S/\varepsilon - 1}{\alpha_{n,1}\tau}$ and $B_{n,1} = (M-1)\log_2(P_{tot}^S/\varepsilon)$, which proves Theorem 2. ∎

*Remarks*: For the CSI accuracy at the BS, $P_{n,1}^p\tau$ (namely transmit energy for training sequence) in TDD systems and $\frac{B_{n,1}}{M-1}$ (namely spatial resolution) in FDD systems are two crucial factors. Specifically, given a requirement on CSI accuracy, it is possible to shorten the length of training sequence by increasing the transmit power, so as to leave more time for data transmission in a time slot. However, in order to keep the pairwise orthogonality of training sequences, the length of training sequence $\tau$ must be larger than the number of MUs. In other words, the minimum value of $\tau$ is $NK$. Similarly, in FDD systems, it is possible to reduce the feedback bits by increasing the number of antennas $M$. Yet, in order to fulfill the spatial degrees of freedom for ZFBF at the BS, $M$ must be not smaller than $(N-1)K+1$. This is because the beam $\mathbf{w}_i$ for the $i$th cluster should be in the null space of the channels for the $(N-1)K$ MUs in the other $N-1$ clusters.

Furthermore, substituting (36) and (37) into (35), we have

$$R_{n,1} \approx -\log_2(1-\rho_{n,1}) + \log_2(\theta_{n,1}) - \sum_{i=1}^{N-1} \Xi_{N-1}\left(i, \{\mu_{n,1}^v\}_{v=1}^{N-1}\right) \log_2\left(\mu_{n,1}^i\right). \tag{38}$$

Given a power allocation scheme, it is interesting that the bound of $R_{n,1}$ is independent of channel conditions. As analyzed above, it is possible to improve the average rate by improving the CSI accuracy. Especially, for FDD systems, we have the following lemma:

*Lemma 3*: At the high power region with a large number of feedback bits, the average rate of the 1st MU increases linearly as the numbers of feedback bits increase.

*Proof:* Replacing $\rho_{n,1}$ in (38) with $\varrho_{n,1} = 1 - 2^{-\frac{B_{n,1}}{M-1}}$, $R_{n,1}$ is transformed as

$$R_{n,1} \approx \frac{B_{n,1}}{M-1} + \log_2(\theta_{n,1}) - \sum_{i=1}^{N-1} \Xi_{N-1}\left(i, \{\mu_{n,1}^v\}_{v=1}^{N-1}\right) \log_2\left(\mu_{n,1}^i\right), \tag{39}$$

which yields Lemma 3. ∎

## B. Noise Limited Case

If the interference term is negligible with respect to the noise term due to a low transmit power, then the SINR $\gamma_{n,k}, \forall n, k$ is reduced as

$$\gamma_{n,k} = \alpha_{n,k}|\mathbf{h}_{n,k}^H \mathbf{w}_n|^2 P_{n,k}^S, \tag{40}$$

which is equivalent to the interference-free case. As discussed earlier, $|\mathbf{h}_{n,k}^H \mathbf{w}_n|^2$ is $\chi^2(2)$ distributed, then the average transmission rate can be computed as

$$\begin{aligned} R_{n,k} &= \int_0^\infty \log_2\left(1 + P_{n,k}^S \alpha_{n,k} x\right) \exp(-x) dx \\ &= -\exp\left(\frac{1}{P_{n,k}^S \alpha_{n,k}}\right) \mathrm{E_i}\left(-\frac{1}{P_{n,k}^S \alpha_{n,k}}\right). \end{aligned} \tag{41}$$

Note that Eq. (41) is independent of the CSI accuracy, thus it is unnecessary to carry out channel estimation or CSI feedback in this scenario. Since both intra-cluster interference and inter-cluster interference are negligible, ZFBF at the BS and SIC at the MUs are not required, and all optimization schemes asymptotically approach the same performance.

TABLE I
PARAMETER TABLE FOR $(\alpha_{n,k}, \rho_{n,k})$, $\forall n \in [1,3]$ AND $k \in [1,2]$.

| n \ k | 1 | 2 |
|---|---|---|
| 1 | (1.00, 0.90) | (0.10, 0.70) |
| 2 | (0.95, 0.85) | (0.20, 0.75) |
| 3 | (0.90, 0.80) | (0.15, 0.80) |

## V. SIMULATION RESULTS

To evaluate the performance of the proposed multiple-antenna NOMA technology, we present several simulation results under different scenarios. For convenience, we set $M = 6$, $N = 3$, $K = 2$, $B_{tot} = 12$, while $\alpha_{n,k}$ and $\rho_{n,k}$ are given in Tab. I for all simulation scenarios without extra specification. In addition, we use SNR (in dB) to represent $10\log_{10} P_{tot}^S$.

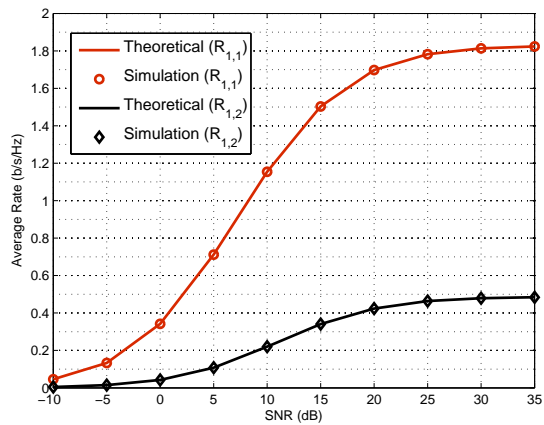

Fig. 2. Comparison of theoretical expressions and simulation results.

First, we verify the accuracy of the derived theoretical expressions. As seen in Fig. 2, the theoretical expressions for both the 1st and the 2nd MUs in the 1st cluster well coincide with the simulation results in the whole SNR region, which confirms the high accuracy. As the principle of NOMA implies, the 1st MU performs better than the second MU. At high SNR, the average rates of the both MUs are asymptotically saturated, which proves Theorem 1 again.

Secondly, we compare the proposed power allocation scheme with the equal power allocation scheme and the fixed power allocation scheme proposed in [4]. Note that the fixed power allocation scheme distributes the power with a fixed ratio 1:4 between the two MUs in a cluster so as to facilitate the SIC. It is found in Fig. 3 that the proposed power allocation scheme offers an obvious performance gain over the two baseline schemes, especially in the medium SNR region. Note that practical communication systems in general operate at medium SNR, thus the proposed scheme is able to achieve a given performance requirement with a lower SNR. As the SNR increases, the proposed scheme and the equal allocation scheme achieve the same saturated sum rate, but the fixed allocation scheme has a clear performance loss.

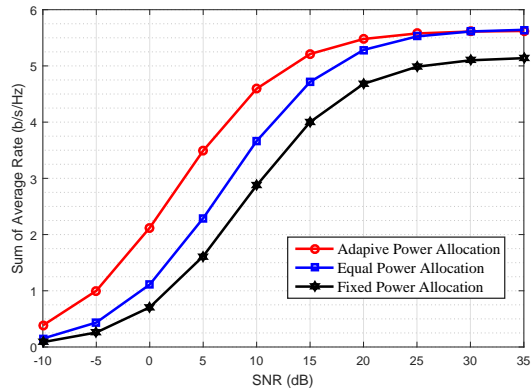

Fig. 3. Performance comparison of different power allocation schemes.

Next, we examine the advantage of feedback allocation for the FDD based NOMA system with equal power allocation, cf. Fig. 4. As analyzed in Section IV.B, at very low SNR, namely the noise-limited case, the average rate is independent of CSI accuracy, and thus the two schemes asymptotically approach the same sum rate. As SNR increases, the proposed feedback allocation scheme achieves a larger performance gain. Similarly, at high SNR, both the two schemes are saturated, and the proposed scheme obtains the largest performance gain. For instance, at SNR= 30 dB, there is a gain of more than 0.5 b/s/Hz. Furthermore, we investigate the impact of the total number of feedback bits on the average rates of different MUs at SNR= 35 dB. As shown in Fig. 5, the performance of the 1st MU is clearly better than that of the 2nd MU. Moreover, the average rate of the 1st MU is nearly a linear function of the number of feedback bits, which reconfirms the claims of Lemma 3.

Then, we investigate the impact of transmission mode on the performance of the NOMA systems at SNR= 10 dB with equal power allocation in Fig. 6. To concentrate on the impact of transmission mode, we set the same CSI accuracy of all downlink channels as $\rho$. Note that we consider four fixed transmission modes under the same channel conditions in the case of 6 MUs in total. Consistent with the claims in Lemma 2, mode 4 with $N=1$ and $K=6$ achieves the largest sum rate at low CSI accuracy, while mode 1 with $N=6$ and $K=1$ performs best at high CSI accuracy. In addition, it is found that at medium CSI accuracy, mode 2 with $N=3$ and $K=2$ is optimal, since it is capable to achieve a best balance between intra-cluster interference and inter-cluster interference. Thus, we propose to dynamically select the transmission mode according to channel conditions and system parameters. As shown by the red line in Fig. 6, dynamic mode selection can always obtain the maximum sum rate.

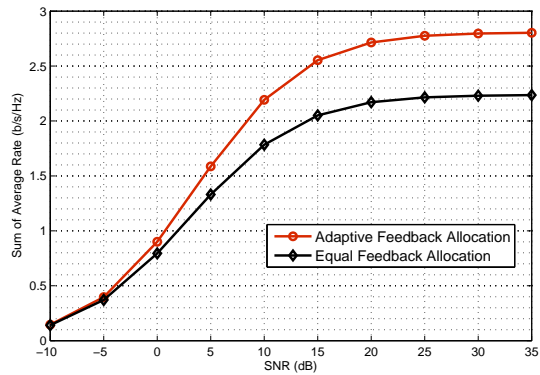

Fig. 4. Performance comparison of different feedback allocation schemes.

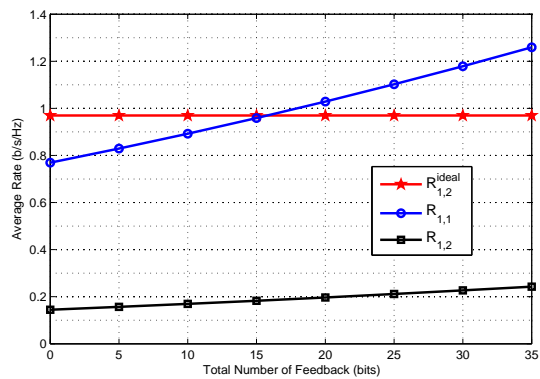

Fig. 5. Asymptotic performance with a large number of feedback bits.

Finally, we exhibit the superiority of the proposed joint optimization scheme for the NOMA

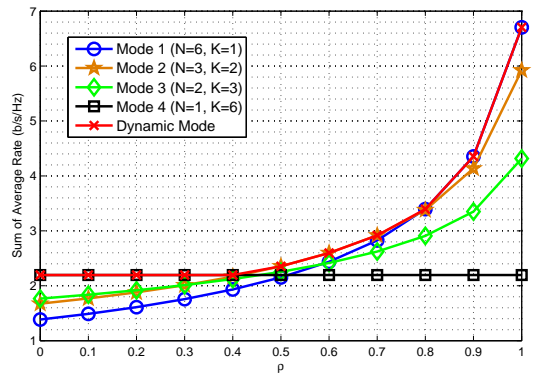

Fig. 6. Performance comparison of different transmission modes.

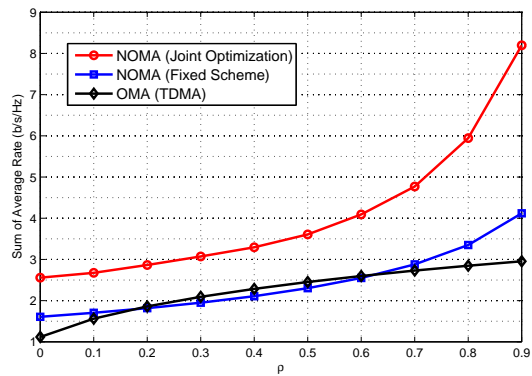

Fig. 7. Performance comparison of a joint optimization scheme and a fixed allocation scheme.

systems at SNR$= 10$ dB. In addition, we take a fixed scheme based on NOMA and a time division multiple access (TDMA) based on OMA as baseline schemes. Specifically, the joint optimization scheme first distributes the transmit power with equal feedback allocation, then allocates the feedback bits based on the distributed power, finally selects the optimal transmission mode. The fixed scheme always adopts the mode 2 ($N = 3, K = 2$) with equal power and feedback allocation. The TDMA equally allocates each time slot to the 6 MUs, and utilizes maximum ratio transmission (MRT) based on the available CSI at the BS to maximize the rate. For clarity of notation, we use $\rho$ to denote the CSI accuracy based on equal feedback allocation. In other words, the total number of feedback bits is equal to $B_{tot} = -K * N * (M - 1) * \log_2(1 - \rho)$. As seen in Fig. 7, the fixed scheme performs better than the TDMA scheme at low and high

CSI accuracy, and slightly worse at medium regime. However, the proposed joint optimization scheme performs much better than the two baseline schemes. Especially at high CSI accuracy, the performance gap becomes substantially large. For instance, there is a performance gain of about 3 b/s/Hz at $\rho = 0.8$, and up to more than 5 b/s/Hz at $\rho = 0.9$. As analyzed in Lemma 2 and confirmed by Fig. 6, when $\rho$ is larger than 0.8, which is a common CSI accuracy in practical systems, mode 2 is optimal for maximizing the system performance. Thus, the joint optimization scheme is reduced to joint power and feedback allocation, which requires only a very low complexity. Thus, the proposed NOMA scheme with joint optimization can achieve a good performance with low complexity, and it is a promising technique for future wireless communication systems.

## VI. Conclusion

This paper provided a comprehensive solution for designing, analyzing, and optimizing a NOMA technology over a general multiuser multiple-antenna downlink in both TDD and FDD modes. First, we proposed a new framework for multiple-antenna NOMA. Then, we analyzed the performance, and derived exact closed-form expressions for average transmission rates. Afterwards, we optimized the three key parameters of multiple-antenna NOMA, i.e., transmit power, feedback bits, and transmission mode. Finally, we conducted asymptotic performance analysis, and obtained insights on system performance and design guidelines.